\documentclass[12pt]{article}
\pdfoutput=1
\usepackage{amsmath}
\usepackage{amssymb}
\usepackage{graphicx}
\usepackage{subcaption}
\usepackage{url}
\usepackage{comment}
\usepackage[utf8x]{inputenc}
\usepackage[sort&compress, numbers, merge]{natbib}

\title{Low Scale GMSB}

\usepackage[colorlinks=true, linkcolor=blue, citecolor=blue,
urlcolor=black]{hyperref}

\begin{document}
\baselineskip 0.6cm

\def\simgt{\mathrel{\lower2.5pt\vbox{\lineskip=0pt\baselineskip=0pt
           \hbox{$>$}\hbox{$\sim$}}}}
\def\simlt{\mathrel{\lower2.5pt\vbox{\lineskip=0pt\baselineskip=0pt
           \hbox{$<$}\hbox{$\sim$}}}}
\def\simprop{\mathrel{\lower3.0pt\vbox{\lineskip=1.0pt\baselineskip=0pt
             \hbox{$\propto$}\hbox{$\sim$}}}}
\def\tr{\mathop{\rm tr}}
\def\SU{\mathop{\rm SU}}

\begin{titlepage}

\begin{flushright}
IPMU17-0171 \\
\end{flushright}

\vskip 1.1cm

\begin{center}

{\Large \bf
Low-scale Gauge Mediation after LHC Run 2
}

\vskip 1.2cm

Xiaochuan Lu${}^{1}$ and Satoshi Shirai${}^{2}$
\vskip 0.5cm

{\it
$^1$ Department of Physics, University of California, Davis, California 95616, USA,
\\
$^2$ {Kavli Institute for the Physics and Mathematics of the Universe (WPI), \\The University of Tokyo Institutes for Advanced Study, The University of Tokyo, Kashiwa
 277-8583, Japan}
}

\vskip 1.0cm

\abstract{We study low-scale gauge mediated supersymmetry breaking models with a very light gravitino of mass $\mathcal{O}(1)$ eV. The cosmological upper bound on the gravitino mass and the collider constraints on the sparticle masses give a significant impact on such models. We apply the latest results of the LHC to these models and obtain the current constraints. We find that perturbatively calculable classes of low-scale gauge mediation models can be largely excluded.}

\end{center}
\end{titlepage}

\section{Introduction}\label{sec:Introduction}

Supersymmetric (SUSY) extensions of the Standard Model (SM) are arguably the most attractive candidates of new physics. Although the mass scale of the SUSY SM (SSM) particles is roughly expected to range from the electroweak scale (100 GeV) to PeV scale, the mass range of the gravitino is quite uncertain. Depending on the SUSY breaking and mediation mechanism, the gravitino mass can vary from eV scale to PeV scale.

Among various possibilities of the gravitino mass, the case of a light ($\mathcal{O}(1)$ eV scale) gravitino is very important, in which the SSM particles can acquire soft masses by the so-called gauge mediated supersymmetry breaking (GMSB)~\cite{Dine:1981za, *Dimopoulos:1981au, *Dine:1981gu, *Nappi:1982hm, *AlvarezGaume:1981wy, *Dimopoulos:1982gm, Dine:1993yw, *Dine:1994vc, *Dine:1995ag} (see also Ref.~\cite{Giudice:1998bp} for a review). This type of models are of great interests for several reasons. First, with such a light gravitino mass, the notorious cosmological gravitino problem on the Big-Bang nucleosynthesis is absent, even with an arbitrarily high reheating temperature. Second, there are no SUSY flavor problems, because the gauge interactions are flavor universal and the contributions from gravity mediation are tiny. Moreover, low-scale mediation models can be described within the framework of quantum field theory, including both the SUSY breaking and mediation sectors, which makes this type of theories calculable and highly predictable.

In GMSB models, the eV-scale gravitino is the lightest SUSY particle (LSP) and stable from decay. However, it cannot be a dominant component of the dark matter, since it is too light to be a cold dark matter. Similar with the case of the neutrino masses, there are cosmological constraints on the gravitino mass due to the warmness of the gravitino dark matter. Previously, the analysis of the Lyman-$\alpha$ forest constrains the gravitino mass as $m_{3/2}<16$ eV ($2\sigma$)~\cite{Viel:2005qj}. Recently this constraint is significantly improved with the CMB lensing and cosmic shear data, giving a new upper bound as $m_{3/2}<4.7$ eV (95\% C.L.)~\cite{Osato:2016ixc}. This translates into stringent upper bounds on the SSM particle masses, which can be confronted at the LHC. In this paper, we discuss the current status of this confrontation after the LHC Run 2.

The minimal types of the GMSB models generically have trouble with explaining the observed value of Higgs mass. As they typically predict a small size of the $A$-term, the scalar top mass needs to be much greater than $\mathcal{O}(1)$ TeV to give a 125 GeV Higgs, which in turn dictates a gravitino mass larger than $\mathcal{O}(10)$ eV~\cite{Ajaib:2012vc}, conflicting the updated upper bound. However, this problem can be solved with slight modifications of the models, such as a singlet extension of the MSSM~\cite{Ellwanger:2009dp, Yanagida:2012ef} and mixing of the MSSM matter field and the messenger field~\cite{Chacko:2001km, Evans:2011bea, *Evans:2012hg, Byakti:2013ti, Evans:2013kxa}, which modify the Higgs sector and enhance the Higgs mass. These modifications are expected to scarcely affect the SUSY signatures at the LHC, as far as they do not significantly affect the nature of the next-to-LSP (NLSP). In this paper, we focus on the direct constraints on the SSM particles, regardless of the Higgs mass.
In the discussion section, we comment on the possible effects on the SUSY signatures for these Higgs sector modifications.

The rest of this paper is organized as follows. In section~\ref{sec:Models}, we discuss the theoretical aspects of low-scale GMSB models, including the typical mass spectra and decay modes of the SSM particles. We then investigate in section~\ref{sec:Collider} the collider constraints on these models, combined with the cosmological bound on the gravitino mass. Section~\ref{sec:Conclusion} is devoted to summaries and some future outlooks.

\section{GMSB Models}\label{sec:Models}

In this section, we discuss the general features of the low-scale GMSB models, with an emphasis on the role of the gravitino mass. We first describe the typical SSM particle spectra in section~\ref{subsec:Upperbound}, and explain why one generically expects upper bounds on the SSM particle masses when the gravitino mass is bounded from above. An estimate of the bound implies that the SSM particles are accessible at the LHC. We then discuss in section~\ref{subsec:Decay} the typical decay modes of the SSM particles, and the resultant collider signatures. In section~\ref{subsec:Setup}, we describe the setup of our own parametrization of benchmark GMSB scenarios.

\subsection{Upper bound on the SSM masses}\label{subsec:Upperbound}

In the GMSB models, SUSY breaking effects are mediated to the SSM sector by the so called messenger particles $\Psi$, which are charged under the SM gauge interactions and coupled to the SUSY breaking sector. The effective interactions between the SUSY breaking sector and the messengers can be expressed by the superpotential
\begin{align}
W \simeq k_{ij} S  \bar{\Psi}_i\Psi_j +  M_{ij} \bar{\Psi}_i\Psi_j  - F^{\dagger}  S,
\label{eqn:Wmess}
\end{align}
where $i,j$ are flavor indices of the messengers, $M_{ij}$ the messenger mass, $S$ the SUSY breaking field, $F$ characterizing the size of the SUSY breaking, and $k_{ij}$ the effective couplings between $S$ and the messenger fields $\Psi_i$. Assuming the scalar component of the superfield $S$ is stabilized at the origin, its vacuum expectation value (VEV) would be $\langle S \rangle = F \theta^2$ and consequently SUSY is broken. The gravitino mass is given as $m_{3/2} = F/\sqrt{3}M_P$, where $M_P$ is the reduced Planck mass.

In hope of a grand unification theory (GUT) of the gauge couplings, we usually assume the messengers to come in complete multiplets of the ${\rm SU}(5)_{\rm GUT}$.
Upon SUSY breaking, the messengers mediate the SUSY breaking effects to the SSM particles, generating soft masses for them through circulating loops. The magnitude of the gaugino mass can be estimated as:
\begin{align}
m_{\rm gaugino} \sim N_5\frac{\alpha}{4\pi} \frac{kF}{M}\left(\frac{kF}{M^2}\right)^p ,
\label{eqn:mgaugino}
\end{align}
with $N_5$ denoting the number of copies of $5+\bar 5$ representations formed by the messengers, $M$ the mass of the messengers, $\alpha$ the corresponding gauge coupling, $k$ the typical size of coupling between the messenger and SUSY breaking sector, and $p$ a model-dependent non-negative parameter. Note that since the GUT symmetry is broken, the messenger mass $M$ and coupling $k$ can be different between the lepton-like and quark-like messengers.
However, once $(k/M)_{\rm quark} = (k/M)_{\rm lepton}$ at the GUT scale, this relation is maintained at lower-energy scale. Therefore, for theories with $p=0$ (as in the case of the minimal GMSB), the GUT relation of gaugino masses $m_{\tilde b}:m_{\tilde w}:m_{\tilde g} = \alpha_1:\alpha_2:\alpha_3$ is maintained down to the messenger scale. For theories with $p>0$, this relation no longer holds~\cite{Nomura:1997uu,Ibe:2007wp,Hamaguchi:2008yu}.

Now let us discuss the upper-bounds on the gaugino masses in light of eq.~\eqref{eqn:mgaugino}. For the messengers not to be tachyonic, one should have $kF/M^2 \lesssim 1$. Consequently, we get the following approximate upper-bounds on the wino and gluino masses:
\begin{align}
m_{\left(\tilde{w},\tilde{g}\right)} \lesssim N_5\frac{\alpha_{(2,3)}}{4\pi} \sqrt{kF} \approx  (2,6)~\text{TeV} \times \sqrt{k} \frac{N_5}{5} \left(\frac{m_{3/2}}{4.7~{\rm eV}}\right)^{1/2} ,
\label{eqn:ubmgaugino}
\end{align}
where the relation $F=\sqrt{3}M_P m_{3/2}$ is used. In the above expression, the various parameters are all bounded from the above. We have $m_{3/2} < 4.7$ eV from the cosmological bound, $N_5\lesssim 5$ to maintain the perturbative grand unification (strict requirement on the perturbative coupling unification leads to even severer constraint as $N_5\leq 4$ \cite{Jones:2008ib}), and $k \lesssim 1$ in case of a perturbative coupling between the SUSY breaking sector and the messengers. All together, they give an upper-bound of the gluino (wino) mass around 6 (2) TeV. As a side remark, we also note that eq.~\eqref{eqn:mgaugino} suggests the messenger mass be of $\mathcal{O}(100)$ TeV in order to give TeV-scale soft masses.

In addition to the above discussions, vacuum (meta-)stability condition typically provides more stringent constraints on the masses of SSM particles. In general, it is difficult to stabilize the scalar component of the SUSY breaking field $S$. For instance, if $k_{ij} \propto \delta_{ij}$ in the superpotential in eq.~\eqref{eqn:Wmess}, there will be a global minimum of the scalar potential, at which SUSY is restored and the messenger fields get VEVs of $\mathcal{O}(\sqrt{F})$. To solve this instability problem, one can make the SUSY breaking vacuum metastable and long lived. A simple way of doing so, for example, is to introduce a non-canonical K\"ahler potential for the SUSY breaking field~\cite{Murayama:2006yf, *Murayama:2007fe}:
\begin{align}
\Delta K = -\frac{1}{4 \Lambda^2} (S^{\dagger} S)^2 + \cdots ,
\end{align}
which corrects the scalar potential as
\begin{equation}
V = \left|\frac{\partial W}{\partial S}\right|^2 \left( 1 + \frac{|S|^2}{\Lambda^2} + \frac{|S|^4}{\Lambda^4} + \cdots \right) ,
\end{equation}
which gives a positive mass term for $S$ at $S=0$ and allows the SUSY broken vacuum to be metastable. Note that $|\Lambda^2| \gg |F|$ is required to justify the picture of effective theory as well as maintain this model perturbatively calculable. To make the lifetime of the metastable vacuum much longer than the age of the Universe, a messenger mass larger than the tachyonic condition $kF\lesssim M^2$ is needed, which suppresses the SSM masses~\cite{Hisano:2007gb, Hisano:2008sy}. In this case, the parameter $p$ in eq.~\eqref{eqn:mgaugino} could be zero and the GUT relation could hold. Hereafter, we refer to this type of spectrum as ``minimal-type" spectrum.

Another way to solve the instability problem is to consider a more complicated type of interactions between the SUSY breaking sector and the messengers to achieve a global SUSY breaking minimum. For instance, specific values of $k_{ij}$ and $M_{ij}$ which satisfy $\mathrm{det}(M+kS)=\mathrm{det}(M)$ and $km^{-1}k=0$ lead to the SUSY breaking global minimum \cite{Komargodski:2009jf,Sato:2009dk}. However, in such cases the parameter $p$ will be positive~\cite{Komargodski:2009jf, Shirai:2010rr}, which further suppresses the gaugino masses. On the other hand, in this kind of solutions, the sfermion masses are generally not suppressed. Therefore, a mass hierarchy between the gaugino and sfermion masses is naturally expected. We refer to this type of spectrum as ``split-type" spectrum.

\subsection{Decay of SSM particles}\label{subsec:Decay}

For collider physics, the decay modes of the produced sparticles are important. In this regards, the eV-scale gravitino plays a crucial role, as it is the LSP. In principle, all the SSM particles can directly decay into the gravitino, with the decay length approximately given by
\begin{equation}
c\tau\simeq 20~{\mu}{\rm m} \left(\frac{m_{3/2}}{1~{\rm eV}} \right)^2\left(\frac{m}{100~{\rm GeV}} \right)^{-5},
\label{eqn:decaykength}
\end{equation}
where $m$ is the mass of the sparticle. However, most sparticles can also decay into lighter sparticles through the SSM interactions, the decay length of which is generically expected to be $\mathcal{O}(\text{GeV}^{-1}) \sim 10^{-16}$ m. Clearly, sparticles would mostly decay into a lighter sparticle instead of gravitino, except for the case of the NLSP, in which there is no lighter sparticle other than the gravitino.

According to the discussion above, once SUSY particles are pair-produced, they will follow a cascade decay procedure into the NLSPs, which eventually decay into the gravitino and SM particles:
$${\rm production} \to {\rm N}^{n}{\rm LSP}\to\cdots \to {\rm NLSP} \to {\rm gravitino}.
$$
As is shown in eq.~\eqref{eqn:decaykength}, the decay of the NLSP is almost prompt in view of the detector resolution for $m_{3/2}=\mathcal{O}(1)$ eV. The gravitinos cannot be directly detected with detectors and are recognized as missing energy (MET). In usual GMSB models, the NLSP is either a neutralino or a slepton. In the former case, the main decay modes of the neutralino NLSP is the gravitino plus a photon, a $Z$ boson, or a Higgs boson. Especially, as events with high-energy photons plus missing energy are very rare in SM processes, picking such signals can significantly reduce the SM background. In the slepton NLSP case, it decays into a gravitino and a lepton. If the produced particles are not sleptons, lepton number conservation requires an additional lepton or neutrino in each SUSY cascade decay. Therefore there are four or more leptons (including neutrinos) in such a SUSY event. This kind of lepton-rich signatures have rather specific features compared to the conventional neutralino LSP scenario~\cite{Asai:2012hb}, and the SM background is drastically reduced.

However, there are caveats to the above discussions, especially in the case of split-type GMSB models. For example, if the bino-like neutralino is the NLSP, it can be as light as $\mathcal{O}(1)$ GeV, without conflicting the collider constraints. In such a case, the decay length of the bino can be much longer than the detector size, which makes it effectively the LSP.

Another issue is the suppression of the decays into the SSM particles, due to heavy intermediate particles. For instance, if the sfermion mass is much larger than the gauginos, the decay rates of the gauginos can be suppressed~\cite{Haber:1983fc, *Baer:1986au, *Barbieri:1987ed, *Baer:1990sc, *Toharia:2005gm, *Gambino:2005eh, Sato:2012xf, *Sato:2013bta}. Approximately, the decay length of the gluino into the bino and wino is
\begin{align}
c\tau_{\widetilde g}(\tilde g \to q\bar{q} \chi) \sim
  10^{-10}~{\rm m}\left(\cfrac{m_{\tilde g}}{1~{\rm TeV}}\right)^{-5}
 \left(\cfrac{ m_{\tilde q}}{10~{\rm TeV}}\right)^{4}.
\end{align}
Comparing to the decay length in eq.~\eqref{eqn:decaykength}, we see that if
\begin{align}
m_{3/2} \ll 1~\text{eV} \left(\cfrac{ m_{\tilde q}}{10~{\rm TeV}}\right)^{2} ,
\end{align}
the gluino dominantly decays into the gravitino. Similar story happens to the wino decaying into bino. When the mass difference between the wino and bino is smaller than the $Z$ boson mass and the Higgsinos are heavy, this decay rate is also suppressed~\cite{Nagata:2015pra, Rolbiecki:2015gsa}, and the wino will directly decay into the gravitino. The branching fraction of the neutral wino into the photon and gravitino is smaller than that of the bino decay, and hence the di-photon signal becomes less strong.

\subsection{Setup of benchmark scenarios}\label{subsec:Setup}

In the framework of the general gauge mediation~\cite{Meade:2008wd}, the gaugino and sfermion soft masses ($M_a$ and $m^2_{\tilde f}$) at the messenger scale are determined by six parameters $\Lambda_{g,a}, \Lambda_{s,a}^2$ with $a=1,2,3$:
\begin{subequations}
\begin{align}
M_{a} &= \frac{\alpha_a}{4\pi} \Lambda_{g,a}, \\
m^{2}_{\tilde f} &= \sum_{a=1}^3 \left(\frac{\alpha_a}{4\pi}\right)^2 C_a(\tilde{f})\Lambda^2_{s,a} ,
\end{align}
\end{subequations}
where $C_a(\tilde{f})$ are the Casimir invariants for the visible particles $\tilde{f}$. Note that the parameter ``$\Lambda_{s,a}^2$'' is a symbol as a whole and can be negative. For convenience we further define ``$\Lambda_{s,a}$'' from it as $\Lambda_{s,a} \equiv \text{sgn}(\Lambda_{s,a}^2) \sqrt{|\Lambda_{s,a}^2|}$. In conventional minimal GMSB model, $\Lambda_{g,a}=N_5\times kF/M$ and $\Lambda_{s,a}^2=N_5 \times 2 k^2 F^2/M^2$.

\begin{figure}[!ht]
\centering
\subcaptionbox{\label{fig:10_mass}$\tan\beta=10$}{
\includegraphics[width=0.47\textwidth]{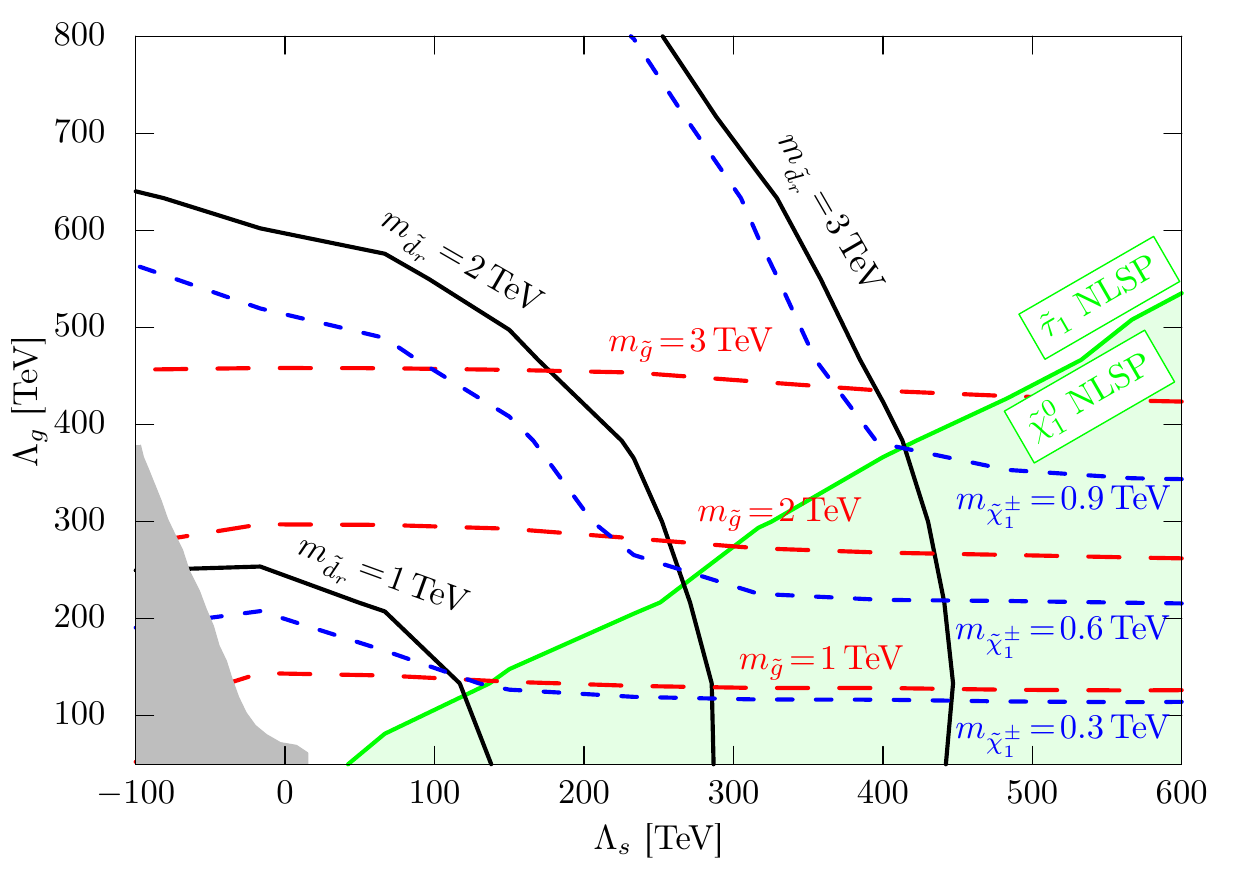}
}
\subcaptionbox{\label{fig:40_mass}$\tan\beta=40$}{
\includegraphics[width=0.47\textwidth]{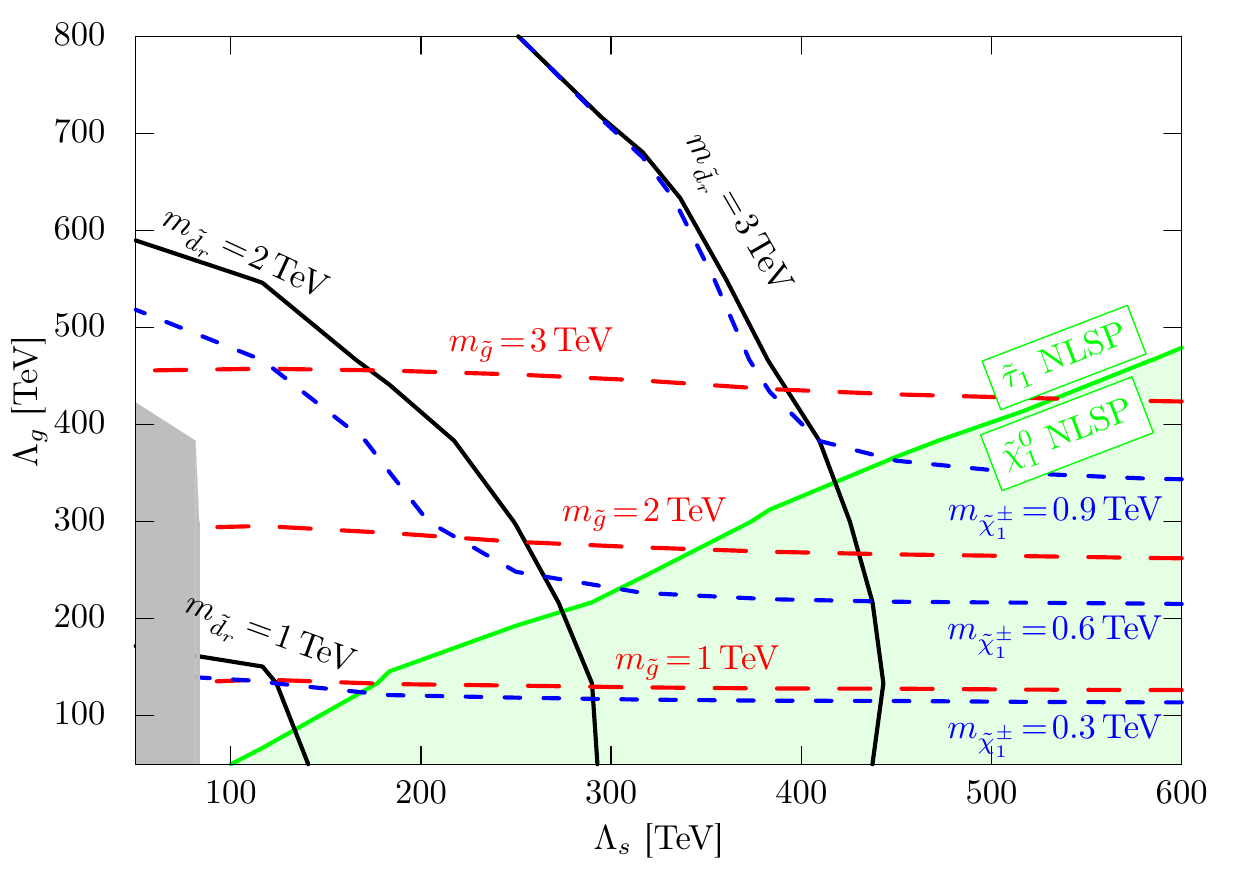}
}
\caption{The masses of sparticles relevant for the dominant production modes in the minimal-type GMSB, with (a) $\tan\beta=10$ and (b) $\tan\beta=40$. In each figure, the region not achieving a correct electroweak symmetry breaking is shaded in gray. The NLSP is a bino-like neutralino in the green shaded region, and the stau elsewhere.}
\label{fig:minimal_mass}
\end{figure}

\begin{figure}[!ht]
\centering
\includegraphics[width=0.94\textwidth]{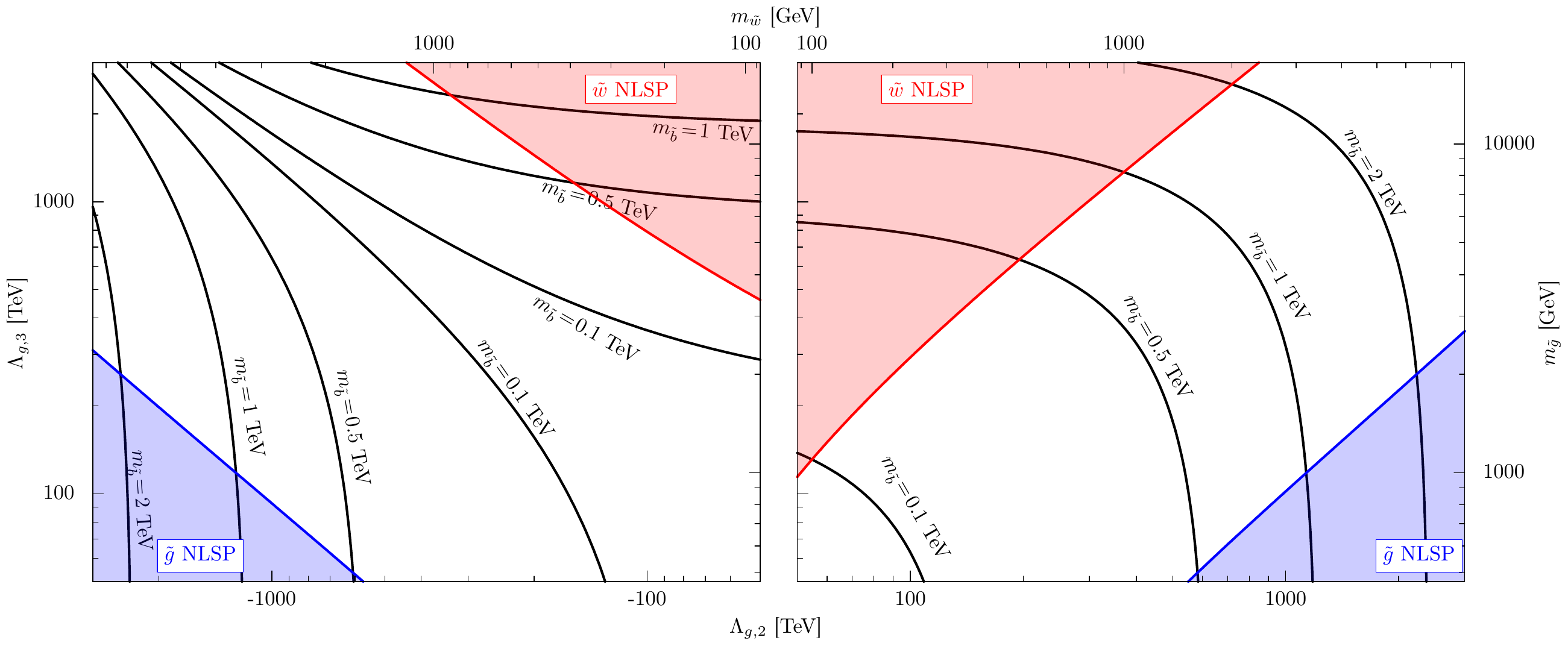}
\caption{The mass spectrum in the split-type GMSB. The region with a gluino (wino) NLSP is shaded in blue (red).}
\label{fig:split_mass}
\end{figure}

In our analysis, we consider two benchmark scenarios, the ``minimal-type'' spectrum and the ``split-type'' spectrum. The first one is motivated by the minimal GMSB model, but slightly extended as in Ref.~\cite{Cheung:2007es}, allowing a general number of messengers. In this case, we assume the GUT relation of the soft masses $\Lambda_{g,a}=\Lambda_g$ and $\Lambda_{s,a} = \Lambda_s$ for $a=1,2,3$, but take $\Lambda_g$ and $\Lambda_s$ as independent free parameters. We fix the messenger mass at $200$ TeV and set the sign of the Higgsino $\mu$-term positive. As the collider signatures depend on the size of $\tan\beta$ (see discussions in section~\ref{subsec:resultsmGMSB}), two benchmark values are considered, $\tan\beta=10$ and $\tan\beta=40$. The masses of sparticles relevant for the dominant production modes (gluino $\tilde g$, squark representative $\tilde d_r$, and lighter charginos $\tilde\chi_1^\pm$) are shown in Fig.~\ref{fig:minimal_mass}. The NLSP in this scenario is either a bino-like neutralino $\tilde\chi_1^0$ or the stau $\tilde\tau_1$.

In the second scenario, we set $\Lambda_{s,a}=\Lambda_s$ (for $a=1,2,3$) large enough such that the sfermions are out of the reach of the LHC (specifically, we choose $\frac{1}{16\pi^2}\Lambda_s=10$ TeV). For the gauginos, we relax the GUT relation into
\begin{equation}
\Lambda_{g,1} = \frac{3}{5}\Lambda_{g,2} + \frac{2}{5}\Lambda_{g,3} ,
\end{equation}
allowing $\Lambda_{g,2}$ and $\Lambda_{g,3}$ to be independent parameters. This can be realized when the messengers belong to the fundamental representation of the $\SU(5)_\text{GUT}$. We take $\tan\beta=1.2$, small enough for a successful electroweak symmetry breaking. The messenger mass is again fixed at $200$ TeV, and the sign of the Higgsino $\mu$-term is set to be positive. The masses of relevant sparticles in the split-type GMSB are shown in Fig.~\ref{fig:split_mass}. In this scenario, the NLSP is mostly bino across the parameter space, but can be wino or gluino (the shaded regions).

\section{Collider Test of GMSB}\label{sec:Collider}

\subsection{Simulation setup}

In our simulations, we set the gravitino mass $m_{3/2}=4$ eV. We compute the mass spectra of SUSY models with SOFTSUSY 3.7.3~\cite{Allanach:2001kg}, and the decay tables with SDECAY 1.3~\cite{Muhlleitner:2003vg}.\footnote{In the stau NLSP regime, for non-NLSP sleptons, SDECAY does not accommodate their direct (two-body) decays into the gravitino, or the three-body decays into the stau mediated by a heavy off-shell neutralino/chargino. As these decay modes can be important for us, we added them by hand on top of the SDECAY output. We followed Ref.~\cite{Ambrosanio:1997bq} in computing the decay widths of off-shell neutralino/chargino mediated three-body decays.} The collider Monte Carlo simulations are then performed with the MG5\_aMC@NLO 2.3.3 event generator~\cite{Alwall:2014hca,*Alwall:2011uj}, interfaced to Pythia 6.4.28~\cite{Sjostrand:2006za} and Delphes 3.3.0~\cite{deFavereau:2013fsa,*Ovyn:2009tx} (with FastJet incorporated~\cite{Cacciari:2011ma,*Cacciari:2005hq}). For colored particle productions, the cross sections are obtained by using NLL-fast v2.1 (for 8TeV) and v3.0 (for 13TeV)~\cite{Beenakker:1996ch,*Kulesza:2008jb,*Kulesza:2009kq,*Beenakker:2009ha,*Beenakker:2011fu,*Beenakker:1997ut,*Beenakker:2010nq}, which are at next-to-leading order in the strong coupling constant, with the resummation of soft gluon emission at next-to-leading-logarithmic accuracy. The parton distribution functions (PDFs) from CTEQ6L1~\cite{Pumplin:2002vw} are used.

\subsection{Relevant experimental searches}

As described in section~\ref{subsec:Decay}, the typical collider signatures of our models are $\gamma\,+$ MET in the neutralino NLSP case, $\tau\,+$ MET and multi-leptons $+$ MET in the slepton NLSP case. In addition, the 0-lepton $+$ multi-jet $+$ MET search might also be constraining, especially in the neutralino NLSP case. Therefore, we include the following ATLAS searches in our analysis: ``Photon8''~\cite{Aad:2015hea}  at the 8TeV run, and ``0lep13''~\cite{ATLAS:2017cjl,*ATLAS:2016kts}, ``23L13''~\cite{ATLAS:2017uun,*ATLAS:2016uwq}, ``Ditau13''~\cite{ATLAS:2017ocr}, ``Photon13''~\cite{ATLAS:2016fks,*ATLASCollaboration:2016wlb}, ``SS3L13''~\cite{Aaboud:2017dmy,*ATLAS:2016kjm}, ``Tau13''~\cite{Aaboud:2016zpr} at the 13TeV run. The luminosities are respectively $20.3~\text{fb}^{-1}$ for Photon8, $3.2~\text{fb}^{-1}$ for Tau13, $13.3~\text{fb}^{-1}$ for Photon13, and $36.1~\text{fb}^{-1}$ for 0lep13, 23L13, Ditau13 and SS3L13. Before briefly describing the cut features of each of these searches in below, we would like to stress that although our own simulation setup is not precisely the same as that used by the ATLAS, we have checked that for each of the searches above, the ATLAS constraints on their chosen benchmark models are well reproduced by us.

\subsubsection*{Photon search}

The photon searches Photon8 and Photon13 look for high $p_\text{T}$ photons together with large MET, which are geared towards the exploration of gauge mediation models with neutralino NLSP. In Photon8~\cite{Aad:2015hea}, there are various signal regions, respectively tailored for a bino-like NLSP scenario under the gluino production (requiring di-photon with high leptonic or jet activities), wino production (requiring di-photon with moderate leptonic or jet activities), and higgsino production (requiring $\gamma + b$ or $\gamma + j$ signatures), as well as a wino-like set of NLSP scenario (requiring $\gamma + l$ signature with low jet activities). On the other hand, the search Photon13 is only sensitive to colored production modes, as it requires large jet activities~\cite{ATLAS:2016fks,*ATLASCollaboration:2016wlb}. Therefore, it is important to keep Photon8 in our analysis for electroweak production modes. In these searches, cuts on the minimum separations of the azimuthal angle $\Delta\phi_\text{min}\left(\text{jet}, p_\text{T}^\text{miss}\right)$ and $\Delta\phi_\text{min}\left(\gamma, p_\text{T}^\text{miss}\right)$ are placed to reduce fake MET from energetic jets or photons. Requirements on the number of jets $N_\text{jets}$, and the variables $H_\text{T}\equiv\sum_{\gamma, l, \text{jets}} p_\text{T}^{}$, $m_\text{eff}\equiv H_\text{T}+E_\text{T}^\text{miss}$, $H_\text{T}^\text{jets}\equiv\sum_\text{jets} p_\text{T}^{}$ are generically used to control the jet activities. In some channels with four or more jets showing up, a cut of $R_\text{T}^4 \equiv \frac{\sum_\text{4 leading jets} p_\text{T}^{}}{\sum_\text{All jets} p_\text{T}^{}}$ not being too close to 1 is also placed to help selecting out events with relatively harder sub-leading jets and hence suppress the SM background.

\subsubsection*{Tau search}

The $\tau$ searches Tau13 and Ditau13 look for hadronically decaying $\tau$ leptons together with large MET. They are potentially important for our stau NLSP regime. In Tau13 search~\cite{Aaboud:2016zpr}, the $1\tau$ signal regions require precisely one $\tau$ candidate and no leptons, while $2\tau$ signal regions require two or more tau candidates with any number of leptons. It demands high jets activities by requiring big $N_\text{jets}$, or high values of the variables $H_\text{T}\equiv\sum_{\tau, l, \text{jets}} p_\text{T}^{}$, $H_\text{T}^\text{2j}\equiv p_\text{T}^{j1} + p_\text{T}^{j2} + \sum_{\tau, l} p_\text{T}^{}$, or $m_\text{eff}\equiv H_\text{T}^\text{2j}+E_\text{T}^\text{miss}$. Therefore, it is not sensitive to electroweak production modes. A minimum separation of $\Delta\phi_\text{min}\left(\text{jet}, p_\text{T}^\text{miss}\right)$ and $\Delta\phi_\text{min}\left(\tau, p_\text{T}^\text{miss}\right)$ is required to reduce fake MET from mismeasured highly energetic jets. Cuts on the transverse masses formed between $\vec{p}\,_\text{T}^\text{miss}$ and the $\tau/l/$jets ($m_\text{T}^\tau$, $m_\text{T}^{\tau_1}+m_\text{T}^{\tau_2}$, $m_\text{T}^\text{sum}\equiv m_\text{T}^{\tau_1}+m_\text{T}^{\tau_2}+m_\text{T}^\text{jets}$, or $m_\text{T}^l$) are used to suppress the $W+$jets background. For certain $2\tau$ regions, the \textit{stransverse} mass variable $m_\text{T2}^{\tau\tau}$~\cite{Lester:1999tx} is used to reduce the $t\bar t$ and $WW$ background. The Ditau13 search is tailored for the productions of charginos and neutralinos~\cite{ATLAS:2017ocr}. It requests at least two quite energetic tau candidates with opposite electrical charge, no $b$-jets nor $\tau$ pairs with invariant mass close to the mass of $Z$ boson. It also uses the $m_\text{T2}^{\tau\tau}$ variable to reduce the background.

\subsubsection*{Lepton search}

There are two important multi-lepton + MET searches for our analysis: 23L13 and SS3L13. The search 23L13~\cite{ATLAS:2017uun,*ATLAS:2016uwq} is geared towards exploring the electroweak production of charginos, neutralinos, and sleptons. It has three signal regions: $2l+0$jets, $2l +$jets, and $3l$. Each region requires exactly two or three leptons respectively in the final state. The $2l+0$jets region targets chargino pair and slepton pair productions. It contains a number of inclusive channels, as well as channels binned by the two leptons' invariant mass $m_{ll}$ and their stransverse mass $m_\text{T2}^{ll}$. Both same flavor and different flavor lepton pairs are allowed. But in case of same flavor lepton pair, their invariant mass is required to be nowhere close to the mass of $Z$ boson. On the other hand, the $2l +$jets region targets $\tilde\chi_1^\pm \tilde\chi_2^0$ productions with decays via SM gauge bosons. So it requests a pair of same-flavor opposite-sign (SFOS) leptons consistent with the $Z$ boson, and at least two jets from the $W$ boson. The $3l$ region targets $\tilde\chi_1^\pm \tilde\chi_2^0$ productions with decays via sleptons or SM gauge bosons into a three-lepton final state. Two of the three leptons are required to form a SFOS pair, the invariant mass of which is required to be consistent (inconsistent) with the $Z$ mass for channels targeting the gauge boson (slepton) mediated decays. Although optimized for electroweak productions, the search 23L13 can be sensitive to the color production modes as well, because the $2l +$jets and $3l$ regions allow for high jet activities. The search SS3L13 requires at least three leptons or a pair of same-sign leptons~\cite{Aaboud:2017dmy,*ATLAS:2016kjm}. The signal regions are classified by the requirements on the minimum number of leptons and the number of $b$-jets, targeting different decay modes of the gluinos and squarks. In addition, multiple energetic jets are also requested. Hence this search is mostly only sensitive to colored particle productions.

\subsubsection*{Energetic jets search}

The search 0lep13 looks for events with no leptons but 2-6 energetic jets and large MET, which is sensitive to the strong production of gluinos and squarks~\cite{ATLAS:2017cjl,*ATLAS:2016kts}. This search is potentially useful to us, especially for the neutralino NLSP regime where the lepton number in the final states is low. There are two types of signal regions, Meff-based and RJR-based, each with various inclusive subregions. For simplicity, only Meff-based regions are included in our analysis, as RJR-based regions give comparable results. We have included more than 20 (overlapping) Meff-based signal regions, defined by different requirements on the minimum number of jets and the cut on $m_\text{eff}\equiv E_\text{T}^\text{miss} + \sum_\text{jets} p_\text{T}^{}$.

\subsection{Results on the minimal-type GMSB}\label{subsec:resultsmGMSB}

In the GMSB models, the NLSP can decay into an energetic lepton or photon, plus missing energy. The SM background for these signatures can be significantly reduced even if one excludes high $p_\text{T}$ jets. Therefore, in addition to the colored sparticle $\tilde g/\tilde q$ productions, the electroweak (EW) productions of the sleptons, charginos and neutralinos can also provide important constraints on the GMSB models. To see this effect, we study the constraints from the strong productions ($\tilde g/\tilde q$) and the electroweak productions ($\tilde\chi/\tilde \ell$) separately.

\begin{figure}[!ht]
\centering
\subcaptionbox{\label{fig:10c}$\tan\beta=10$, $\tilde{g}/\tilde{q}$ production}{
\includegraphics[width=0.47\textwidth]{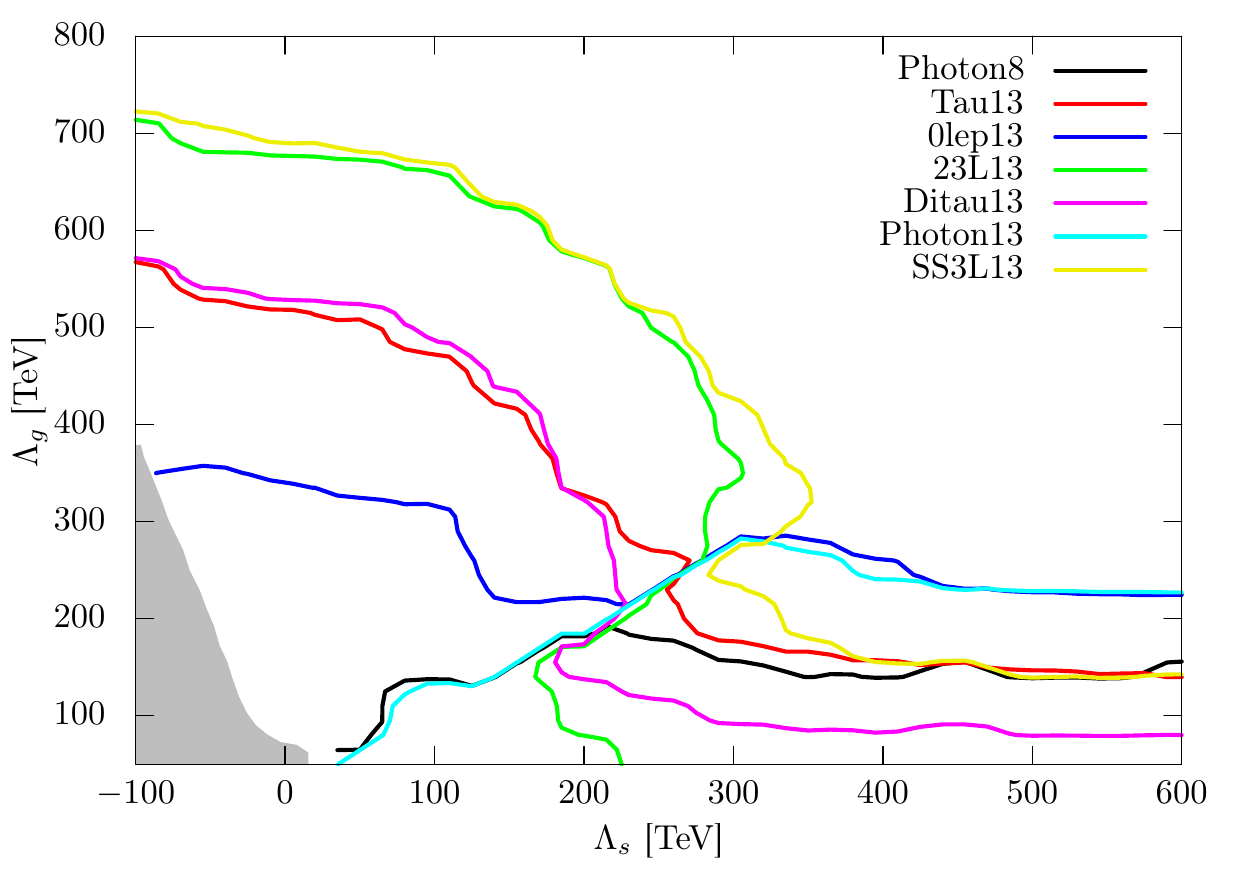}
}
\subcaptionbox{\label{fig:10nc} $\tan\beta=10$, $\tilde{\chi}/\tilde{\ell}$ production}{\includegraphics[width=0.47\textwidth]{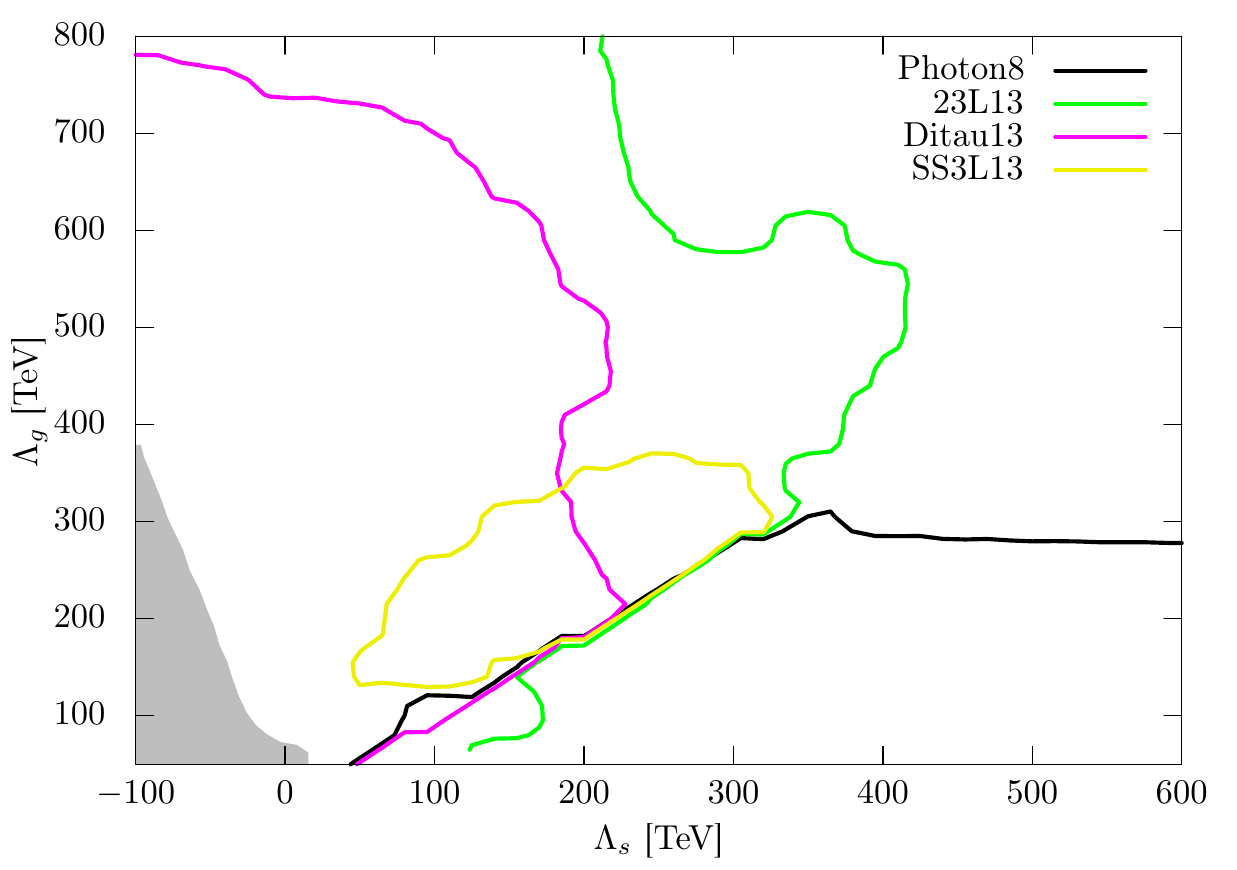}
}
\subcaptionbox{\label{fig:40c}$\tan\beta=40$, $\tilde{g}/\tilde{q}$ production}{
\includegraphics[width=0.47\textwidth]{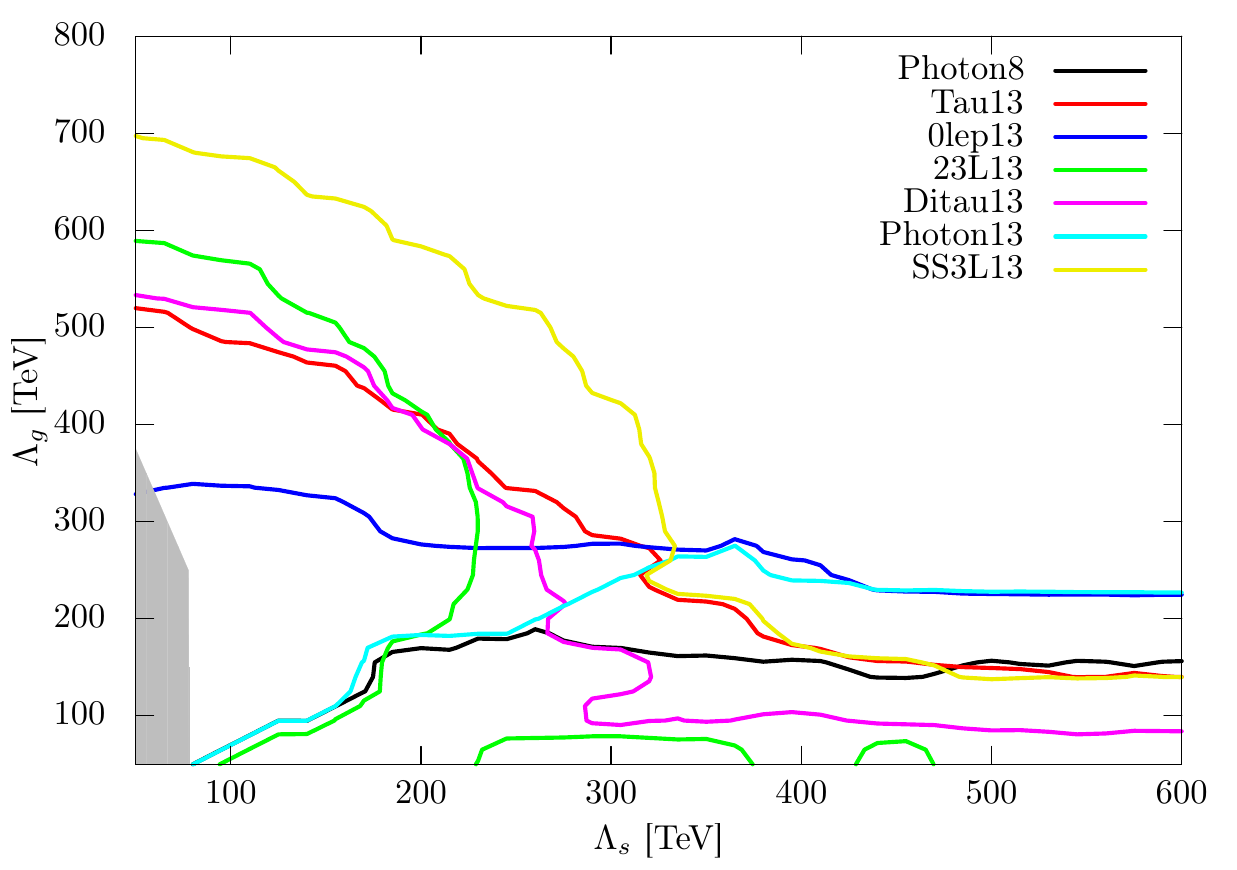}
}
\subcaptionbox{\label{fig:40nc} $\tan\beta=40$,  $\tilde{\chi}/\tilde{\ell}$  production}{\includegraphics[width=0.47\textwidth]{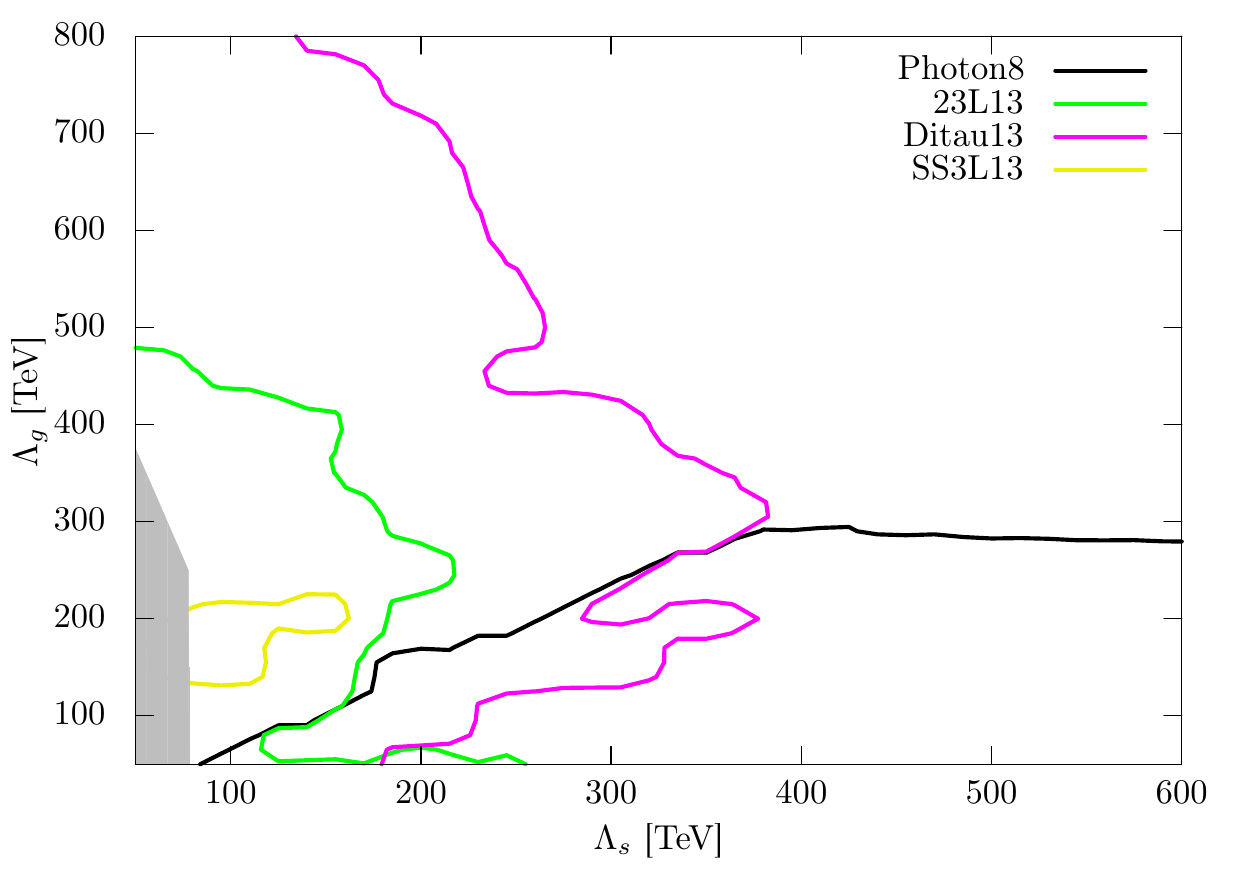}
}
\caption{
The exclusion limits on the minimal-type GMSB parameter space from each LHC search.
}
\label{fig:minimal_individual}
\end{figure}

In Fig.~\ref{fig:minimal_individual}, we show the constraints on the minimal-type GMSB models from each LHC search, with four panels showing different cases ($\tan\beta=10$ or $\tan\beta=40$, and $\tilde g/\tilde q$ or $\tilde\chi/\tilde\ell$ productions). Depending on the size of $\tan\beta$, the LHC signatures are different in the slepton NLSP regime. If $\tan\beta$ is small enough, the selectron, smuon and stau have almost the same mass. In this co-NLSP case, the selectron and smuon can directly decay into the gravitino, emitting energetic electron and muon. But as $\tan\beta$ gets larger, the stau mass gets smaller and all the cascade decays end at the stau, emitting a tau lepton~\cite{Ambrosanio:1997bq, Hamaguchi:2007ge,*Hamaguchi:2007ji}. Therefore, in the case of large $\tan\beta$, the LHC event is tau-lepton rich. In the bino NLSP regime, photon signatures strongly constrain the parameter space.

\begin{figure}[!ht]
\centering
\subcaptionbox{\label{fig:10all}$\tan\beta=10$}{
\includegraphics[width=0.47\textwidth]{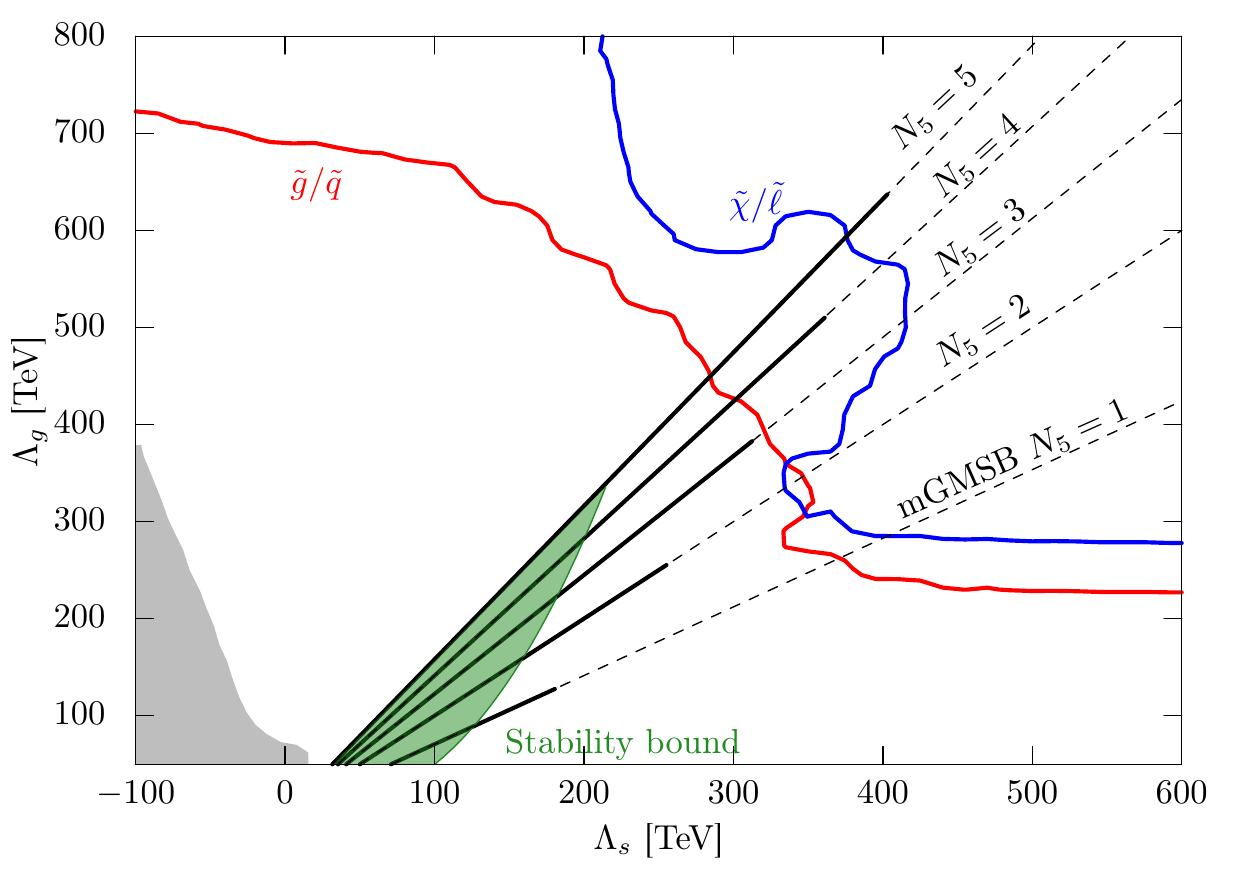}
}
\subcaptionbox{\label{fig:40all} $\tan\beta=40$}{\includegraphics[width=0.47\textwidth]{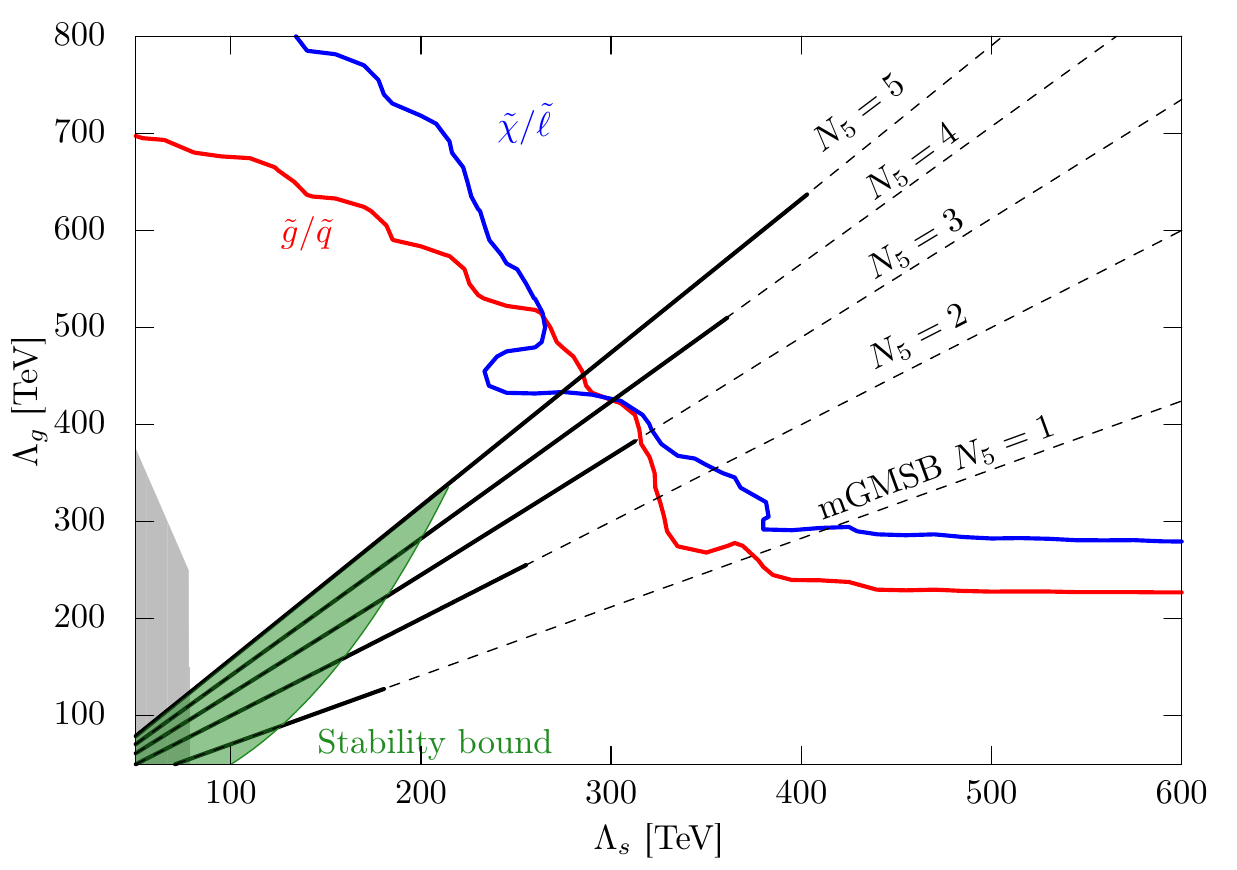}
}
\caption{
The combined exclusion limits on the minimal-type GMSB models.
The black lines of ``mGMSB" show slope of $(\Lambda_s, \Lambda_g) = (\sqrt{2N_5}k F/M ,N_5 k F/M)$ as in the conventional minimal GMSB models.
The solid black lines show the range of $k<1$ and $m_{3/2}<4.7$ eV.
The green shaded regions show the range of the metastability limits at zero temperature.
}
\label{fig:minimal_combined}
\end{figure}

A combine of the various constraints from different searches, according to best expected sensitivity, is shown in Fig.~\ref{fig:minimal_combined}, where we have added some theoretical implications. The black lines (solid and dashed) represent the minimal gauge mediation models with the number of messengers $N_5=1,2,\cdots,5$ respectively. The solid part shows the range $\Lambda_g < N_5\times 130$ TeV, which approximately corresponds to the maximal value under $k<1$ and $m_{3/2}<4.7$ eV. The green region show the metastability bound on the GMSB model, assuming $\Lambda/\sqrt{F}=2$ \cite{Hisano:2007gb}. If we further assume stability of the vacuum at high temperature in the early Universe, the constraints will get much severer~\cite{Hisano:2008sy}. We see that electroweak production of the sparticles more effectively constrain the parameter space. The region that perturbative vacuum stability is achievable is strongly disfavored.

In the strong production modes, roughly speaking gluino and squark productions are equally important. They have comparable cross sections and event acceptance rates. On the other hand, for the electroweak production modes, the dominate mode varies across the parameter space. In the lower right part of the plots, wino productions are dominantly important. However, as one moves towards the upper left region, the Higgsino $\mu$-term becomes smaller than the wino mass  $m_{\tilde w}$ (as well as the bino mass $m_{\tilde b}$ eventually). As a result, the dominant production mode switches from wino to Higgsino productions. This causes, for example, the turn-over of the 23L13 search constraint in Fig.~\ref{fig:10nc}. If one further goes up towards the very top part of the parameter region, the winos ($m_{\tilde w}$) become too heavy to maintain a sizable cross section, and the major production modes will be Higgsinos and sleptons. In the $\tan\beta=10$ case, the slepton production will give a dominant constraint as the 23L13 search is very effective.

\subsection{Results on the split-type GMSB}

Let us now discuss the constraints on the gaugino masses in the split-type GMSB models. The LHC signatures strongly depend on the type of the NLSP.

\subsubsection*{Bino NLSP}

As we see from Fig.~\ref{fig:split_mass}, the bino is the NLSP for most of the parameter space. In split-type GMSB models, the sfermions and Higgsinos are heavy. Consequently, the direct bino pair production is negligible, and the bino primarily comes from the decays of heavier gauginos.

In some parameter regions, the bino can be very light (see Fig.~\ref{fig:split_mass}) and hence has a large decay length. The displaced bino decay will provide non-prompt photon signatures. As discussed in section~\ref{subsec:Decay}, if the decay length of the bino NLSP is much longer than the detector size, the signatures are very similar to the conventional bino LSP scenario. In such cases, the constraint on the gluino mass is still severe ($|\Lambda_{g,3}|\gtrsim 200$ TeV), but the wino constraint may be relaxed, depending on the decay of the neutral wino. If the main decay mode of the neutral wino is the Higgs boson, the wino constraint is $|\Lambda_{g,2}| \gtrsim 100$ TeV~\cite{Aad:2015jqa}. In the following, we omit this possibility and assume the bino decay is prompt for simplicity.

In the case of the wino production, the prominent LHC signature is $\gamma +$ MET. The search Photon8 is most sensitive and roughly gives a bound $\left|\Lambda_{g,2}\right|\gtrsim 200$ TeV. If the neutral wino mainly decays into a $Z+$bino, the search 23L13 can give a constraint, which is still weaker than the search Photon8. In the case of the gluino production, the LHC signatures are $\gamma +$ MET or 0-lepton + multi-jets + MET. In this case, the search Photon13 is slightly stronger than the search 0lep13 and the constraint is approximately $|\Lambda_{g,3}|\gtrsim 200$ TeV.

\subsubsection*{Wino NLSP}

If the wino is the NLSP, the LHC signatures depend on the mass difference between the charged and neutral winos. For pure winos, the charged component is slightly heavier, and the mass difference is around 170 MeV. Thus the charged wino can decay into the neutral wino emitting a pion, with a decay length around $(5~\text{cm})^{-1}$. This decay width is typically smaller than that into the gravitino and a $W$ boson. So all the winos will dominantly directly decay into the gravitino. Because there is no neutral wino pair production at the tree-level, the di-photon signature is less effective. In this case, we find the constraints on the wino mass around 370 GeV (Photon8) and 500 GeV (23L13).

\subsubsection*{Gluino NLSP}

If the gluino is the NLSP, or if its decay rates into the lighter gauginos are suppressed by large sfermion masses and/or the smallness of the mass difference, the gluino will directly decay into a gluon and the gravitino. In this case, the LHC signatures are jets plus missing energy. We find a lower bound on the gluino mass around $1.9$ TeV (0lep13).

\subsubsection*{Summary}

\begin{figure}[!ht]
\centering
\includegraphics[width=0.94\textwidth]{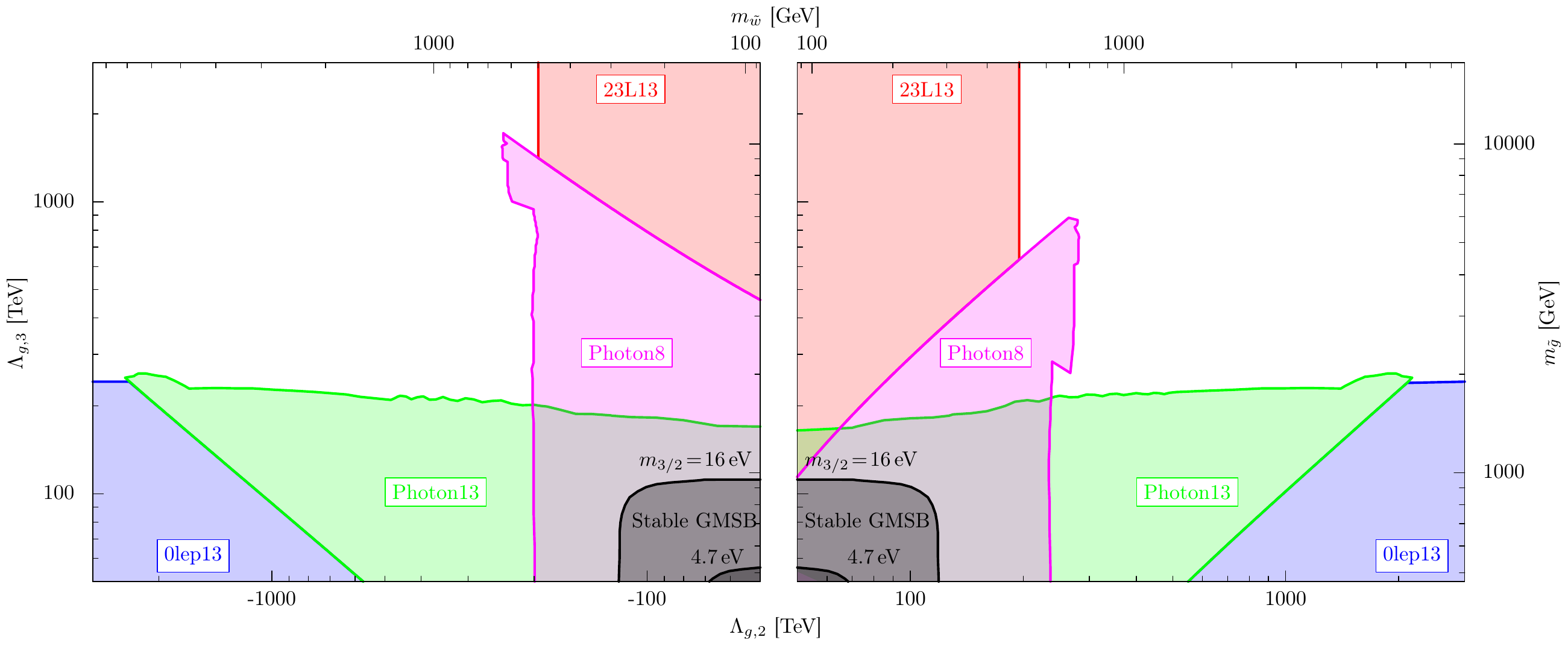}
\caption{The LHC constraints on the split-type GMSB. Shaded regions are excluded by the searches labeled. The gray regions show the bound from the model in Ref.~\cite{Sato:2009dk} with $m_{3/2}=$ 4.7 and 16 eV.}
\label{fig:split_comb}
\end{figure}

In Fig.~\ref{fig:split_comb}, we summarize the current LHC constraints on the split GMSB models. Excluded regions are shaded. We also show the bound from a GMSB model with perturbatively stable vacuum with $m_{3/2}=$ 4.7 and 16 eV~\cite{Sato:2009dk,Sato:2010tz}. The current LHC constraints exclude both cases.

\section{Conclusion and discussion}\label{sec:Conclusion}

We have shown that the present LHC constraints on the SSM particles, together with the cosmological constraint on the gravitino mass, are of critical importance for the low-scale gauge mediation models. Especially, the perturbatively calculable models with stable SUSY breaking vacua are strongly disfavored by these constraints. If we do not restrict to perturbatively calculable models, we can use additional sector~\cite{Hook:2015tra}, and/or strongly interacting messenger sector~\cite{Ibe:2016kyg} to stabilize the SUSY breaking vacua. Although the precise calculation of the SSM mass spectrum is difficult for the strongly interacting models, adopting $k=4\pi$ with eq.~\eqref{eqn:ubmgaugino} can provide rough estimates, and one naively expects the mass bound to be increased by a factor of $\sqrt{4\pi}$. Thus, strongly interacting GMSB models still endure with the present bound on the gravitino mass from cosmology and SSM sparticles masses from the LHC.

There are obviously two directions for a more comprehensive investigation in future. One is to further tighten the cosmological constraint on the gravitino mass. A gravitino mass $\sim 1$ eV can be hopefully tested in near future~\cite{Kamada:2013sya, Oyama:2016lor}, which can further reduce the upper bounds on the SSM sparticles masses by a factor of 2. The other direction is to increase the lower bounds of the SSM masses at future colliders. For this, we roughly estimate the constraints on the masses of the gluino, wino, and higgsino. Very crudely speaking, the current LHC constraints (Fig.~\ref{fig:minimal_combined} and Fig.~\ref{fig:split_comb}) correspond to the following lower bounds on these masses
\begin{align}
m_{\tilde g} \gtrsim 2~\text{TeV} , \quad
m_{\tilde W} \gtrsim 0.7~\text{TeV} , \quad
m_{\tilde h} \gtrsim 0.6~\text{TeV} , \nonumber
\end{align}
which yield the production cross sections as $1$ fb, $14$ fb, and $15$ fb respectively at the $13$ TeV LHC. As the SM background for our signatures is typically quite small, we can estimate future reaches by maintaining the same amount of signal events. For example, at the future LHC with a high luminosity of 3000~$\text{fb}^{-1}$, we find the projected mass bounds can get up to around
\begin{align}
m_{\tilde g} \gtrsim 3~\text{TeV} , \quad
m_{\tilde W} \gtrsim 1.4~\text{TeV} , \quad
m_{\tilde h} \gtrsim 1.3~\text{TeV} . \nonumber
\end{align}
For future 33 TeV LHC with a luminosity of 3000~$\text{fb}^{-1}$, these bounds can even get up to
\begin{align}
m_{\tilde g} \gtrsim 6~\text{TeV} , \quad
m_{\tilde W} \gtrsim 2.5~\text{TeV} , \quad
m_{\tilde h} \gtrsim 2.2~\text{TeV} . \nonumber
\end{align}

Finally, let us comment on the Higgs mass. In this paper, we do not consider constraints from the Higgs mass and conservatively discuss the direct LHC bounds on the SSM sparticles. But let us comment on the case of model modifications to accommodate the Higgs mass. Many extensions of the GMSB models modify the Higgs sector to solve the $b/\mu$ problem and/or enhance the Higgs boson mass. Such extensions will mainly affect the LHC signatures through the change of the Higgsino mass and nature of the NLSP. As seen in Fig.~\ref{fig:minimal_combined}, the electroweak production of the neutralinos/charginos play the most important role in constraining the models. Especially, in the region of the slepton NLSP, the main production modes come from Higgsino-like neutralinos/charginos. Therefore, the exclusion curves will be most likely affected by the change of the Higgsino mass. However, the constraints from the colored sparticle productions will not change much, because the masses of the gluino and 1st/2nd generation squarks will be barely affected, as they couple to the Higgs sector very weakly. Another possible effect is the change in the nature of the NLSP.
The LHC signature strongly depends on the nature of the NLSP \cite{Ruderman:2011vv, Kats:2011qh,Kim:2017pvm}.
For example, in the NMSSM, the singlet-like neutralino can be the NLSP. If it is heavy enough, it mainly decays into the Higgs and $Z$ boson, which leads to less photon signatures compared to the bino NLSP case. However, such $Z$ and Higgs rich events are also strongly constrained by the LHC~\cite{Sirunyan:2017obz}. So generically, we still expect severe constraints, even if the nature of the NLSP is modified.

In the low-scale GMSB models, the LSP is the almost massless gravitino. As far as the produced particles decay into the gravitino inside the detector, either high energy jets, leptons, or photons will be produced accompanied by large missing energy, except in special cases~\cite{Fan:2011yu}. Therefore the low-scale GMSB models are strongly constrained by the LHC.

\section*{Acknowledgments}

SS thanks A. Kamada and K. Yonekura for useful discussion. This work is supported by Grant-in-Aid for Scientific Research from the Ministry of Education, Culture, Sports, Science, and Technology (MEXT), Japan, No. 17H02878 (SS) and by World Premier International Research Center Initiative (WPI), MEXT, Japan (SS). XL is supported by DOE grant DE-SC0009999.


\bibliographystyle{aps}
\bibliography{ref}

\end{document}